\documentclass[preprint,showpacs,superscriptaddress,aps]{revtex4-1}
\usepackage{amssymb}
\usepackage{amsmath,amssymb,graphicx}
\usepackage{graphicx}
\usepackage{dcolumn}
\usepackage{bm}
\def\be{\begin{equation}}
\def\ee{\end{equation}}
\def\bea{\begin{eqnarray}}
\def\eea{\end{eqnarray}}

\def\lsim{\raise0.3ex\hbox{$\;<$\kern-0.75em\raise-1.1ex\hbox{$\sim\;$}}}
\def\gsim{\raise0.3ex\hbox{$\;>$\kern-0.75em\raise-1.1ex\hbox{$\sim\;$}}}

\usepackage{pstricks}

\begin{document}

\title{Direct CP violation in $D^+ \to  K^0(\bar K^0) \pi^+$  decays as a probe for new physics}

\author{David Delepine}
\email{delepine@fisica.ugto.mx}
\affiliation{{\fontsize{10}{10}\selectfont{Division de Ciencias e
Ingenier\'ias,  Universidad de Guanajuato, C.P. 37150, Le\'on,
Guanajuato, M\'exico.}}}

\author{Gaber Faisel}
\email{gaberfaisel@sdu.edu.tr}

\affiliation{{\fontsize{10}{10}\selectfont{Department of Physics,
Faculty of Arts and Sciences, S\"uleyman Demirel University,
Isparta, Turkey 32260.}}}

\author{ Carlos A. Ramirez}
\email{jpjdramirez@yahoo.com}
\affiliation{{\fontsize{10}{10}\selectfont{Depto. de F\'isica,
Universidad de los Andes, A. A. 4976-12340, Bogot\'a, Colombia.}}}

\begin{center}
\date{\today}
\begin{abstract}

In this paper we investigate CP violation in charged decays of $D$
meson. Particularly, we study the direct CP asymmetry of the
Cabibbo favored non-leptonic $D^+ \rightarrow \bar K^0 \pi^+$ and
the doubly Cabibbo-suppressed decay mode $D^+ \rightarrow K^0
\pi^+$ within standard model, two Higgs doublet model with generic
Yukawa structure and left right symmetric models. In the standard
model, we first derive the contributions from box and di-penguin
diagrams contributing to their amplitudes which are relevant to
the generation of the weak phases essential for non-vanishing
direct CP violation. Then, we show that these phases are so tiny
leading to a direct CP asymmetry of order $10^{-11}$ in both decay
modes. Regarding the two Higgs doublet model with generic Yukawa
structure and after taking into account all constraints on the
parameter space of the model, we show that the enhanced direct CP
asymmetries can be 6 and 7 orders of magnitudes larger than the
standard model prediction for $D^+ \rightarrow \bar K^0 \pi^+$ and
$D^+ \rightarrow K^0 \pi^+$ respectively. Finally, within left
right symmetric models, we find that sizable direct CP asymmetry
of ${\mathcal O } (10^{-3})$ can be obtained for the decay mode
$D^+ \rightarrow \bar K^0 \pi^+$ after respecting all relevant
constraints.

\end{abstract}
\end{center}

\maketitle
\flushbottom

\section{Introduction}
Heavy meson decays can serve as a probe for New Physics (NP)
beyond the Standard Model (SM). Of particular interest, CP
violation in heavy mesons decays can discriminate between many
extensions of beyond SM physics that have new complex couplings of
the new particles to quarks or leptons. These couplings provide
the sources of the so called weak phases which are essential for
having non vanishing CP violation. In the SM, complex couplings
can arise only in the Cabibbo-Kobayashi-Maskawa (CKM) matrix
describing the quark mixing
\cite{Cabibbo:1963yz,Kobayashi:1973fv}. The couplings of the
interactions of the charged quarks to $W^\pm$ gauge bosons are
proportional to the CKM matrix elements. Thus, with the presence
of such interactions, CP violation can be generated in the SM.
However, the CP violation in the SM is too small to account for
the observed baryon asymmetry which play an important role in the
domination of matter in our local regions in the universe.

In the mesons sector, CP violation has been observed in the kaon
and B mesons
\cite{Christenson:1964fg,Aubert:2004qm,Aaij:2013iua,Aaij:2012kz}.
Regarding D mesons, the $D^0 -\bar D^0$ mixing was discovered in
2007 after combining the results from BABAR \cite{Aubert:2007wf},
Belle \cite{Staric:2007dt} and CDF \cite{Aaltonen:2007ac}. Later,
the mixing has been observed at LHCb\cite{Aaij:2012nva} and at
Belle \cite{Ko:2014qvu}. Concerning direct CP violation in $D$
meson decays, search in $D^0\to K^+ K^-$ and $D^0\to \pi^+ \pi^-$
has been carried at LHCb \cite{Aaij:2011in,Aaij:2014gsa}, Fermilab
\cite{Collaboration:2012qw} and Belle \cite{Ko:2012px}. Recent
search at LHCb with sensitivities that have reached a level of
$10^{-3}$ has shown that $A_{CP}(D^0\to K^+K^-) = (0.04\pm 0.12
\pm 0.10)\%$ and $A_{CP}(D^0\to \pi^+\pi^-) = (0.07 \pm 0.14 \pm
0.11)\%.$ \cite{Aaij:2016dfb}. Here, $A_{CP}$ stands for the
time-integrated CP asymmetry and clearly, the results show that no
sign of direct CP violation  in these decay modes.

Two body non-leptonic D decays can be sorted into Cabibbo-Favored
(CF), singly Cabibbo-suppressed (SCS) and Double Cabibbo
Suppressed (DCS) according to the suppression factor $\lambda
\simeq |V_{us}| \simeq |V_{cd}|$ appears in their amplitudes. In
the SM, previous studies showed that direct CP-asymmetry of order
$10^{-3}$ can be obtained for some  SCS decay modes
\cite{Bhattacharya:2012ah, Nierste:2017cua}. For examples, the CP
asymmetries of the decays $D^0 \to K_s K^{*^0}$ and $D^0 \to K_s
\bar K^{*^0}$ have been estimated to be as large as $ 3 \times
10^{-3}$ \cite{Nierste:2017cua}. With more investigations in SCS
decay modes, within SM also, a large CP-asymmetry of order
$10^{-2}$ has been predicted for the mode $D^0 \to K_s K_s$
\cite{Nierste:2015zra}. Turning to the CF and DCS two body D
decays, the asymmetries, within SM, are expected to be so tiny and
of order $\lesssim 10^{-9}$ for $D^0 \to K^- \pi^+ $ and $D^0 \to
K^+ \pi^- $ as estimated in our earlier studies in
Refs.\cite{Delepine:2012xw,Delepine:2017oor}. The result motivated
us to explore, also in the same studies,  NP effects in these
decay processes where we have shown that in some extensions of the
SM sizable CP asymmetry of order $10^{-2}$ can be obtained. In
this work, we extend our studies in
Refs.\cite{Delepine:2012xw,Delepine:2017oor} to explore direct CP
violation in charged $D$ decays to CF and DCS $K\pi$ final states.
In particular, we consider the CF mode $D^+ \rightarrow \bar K^0
\pi^+$ and the DCS $D^+ \rightarrow K^0 \pi^+$ decay mode. The
direct CP asymmetries of $D^+ \rightarrow K^0(\bar K^0) \pi^+$ are
expected to be different than those of $D^0 \rightarrow K^\pm
\pi^\mp$ as the strong CP violating phases contributing to these
processes have different origins. In our work in Ref.
\cite{Delepine:2012xw} we found that sizeable direct CP asymmetry
of $D^0 \rightarrow K^- \pi^+$ can be generated in a specific new
physics model namely, in no-manifest Left-Right Symmetric (LRS)
model. In this study, we inspect if the model can still lead to
sizeable CP asymmetries after taking into account the up to date
constraints from collider physics, flavor physics, and low-energy
precision measurements.

 This paper is organized as follows:  in section \ref{dkp}, we
derive the amplitudes of the $\rm CF$ and $\rm DCS$ non leptonic
$D^+ \to  K^0(\bar K^0)  \pi^+$ decays in the framework of the SM
and give our estimations of their direct CP asymmetries. Motivated
by the almost null values of the asymmetries, we extend the
analysis to include two possible candidates of NP models.  These
NP candidates are based on the presence of new charged scalars as
in general two Higgs models in section \ref{Hig} and new charged
bosons as in no-manifest LRS in section \ref{LRSS}. Finally, we
conclude in sect. \ref{CONC}.

\section{ Direct CP asymmetry of  $\rm CF$ and $\rm
DCS$ non leptonic $D^+ \to K \pi^+$ decays in the standard model
\label{dkp}}

In general the effective Lagrangian describing   $\rm CF$ and $\rm
DCS$ $D^+ \to K \pi^+$ decays can be expressed as

\begin{eqnarray}
{\cal L}_{\rm eff.} &=& {G_F\over\sqrt{2}}V_{c q}^*V_{u q'}
\left[\sum_{i,\ a} c_{1ab}^i \big(\bar q \, \Gamma^i \, c_a \big)
\big(\bar u \, \Gamma_i \, q'_b\big) + \sum_{i,\ a} c_{2ab}^i
\big(\bar u \, \Gamma^ic_a \big)\, \big(\bar q \, \Gamma_i \, q'_b
\big)\right]\label{ef}
\end{eqnarray}

Here $i=$ S, V and T stands for scalar (S), vectorial (V) and
tensorial (T) operators respectively. The Latin indexes $a,\ b=L,\
R$ and $q'_{L,\ R}=(1\mp \gamma_5)q$. In Eq. (\ref{ef}) we have
$q\neq q'$ where $q$ and $q'$ can be $d$ or $s$ down-type quark.
For $\rm CF$ decays  $q =s $ and  $q' = d$ while for $\rm DCS$
decays we have $q =d $ and  $q' = s$.

In the SM the contributions from tree-level and loop-level
diagrams, shown in Fig., lead to the effective Hamiltonian that
can be expressed as

\begin{eqnarray}
{\cal H}^{SM}_{\rm eff.} &=& {G_F\over\sqrt{2}}V_{c q}^*V_{u
q'}\bigg[c_1 \big( \bar q \gamma_\mu c_L\big) \big( \bar u
\gamma^\mu q'_L\big)
+\big( c_2\bar u \gamma_\mu c_L \big) \big(\bar q \gamma^\mu q'_L\big) \bigg]+{\rm h.c.}\nonumber \\
&=& {G_F\over\sqrt{2}}V_{c q}^*V_{u q'}\left(c_1\, Q_1+c_2\,
Q_2\right)+{\rm h.c.} \label{SMH}
\end{eqnarray}

In the framework of naive factorization approximation (NFA), the
amplitude of a given decay process under concern can be  obtained
using ${\cal H}^{SM}_{\rm eff.}$ via

\be A^{SM}_{D^+\to K \pi^+} = \langle  K \pi^+ | {\cal
H}^{SM}_{\rm eff.} | D^+ \rangle  \ee

\begin{figure}[tbhp]
\includegraphics[width=8cm,height=4cm]{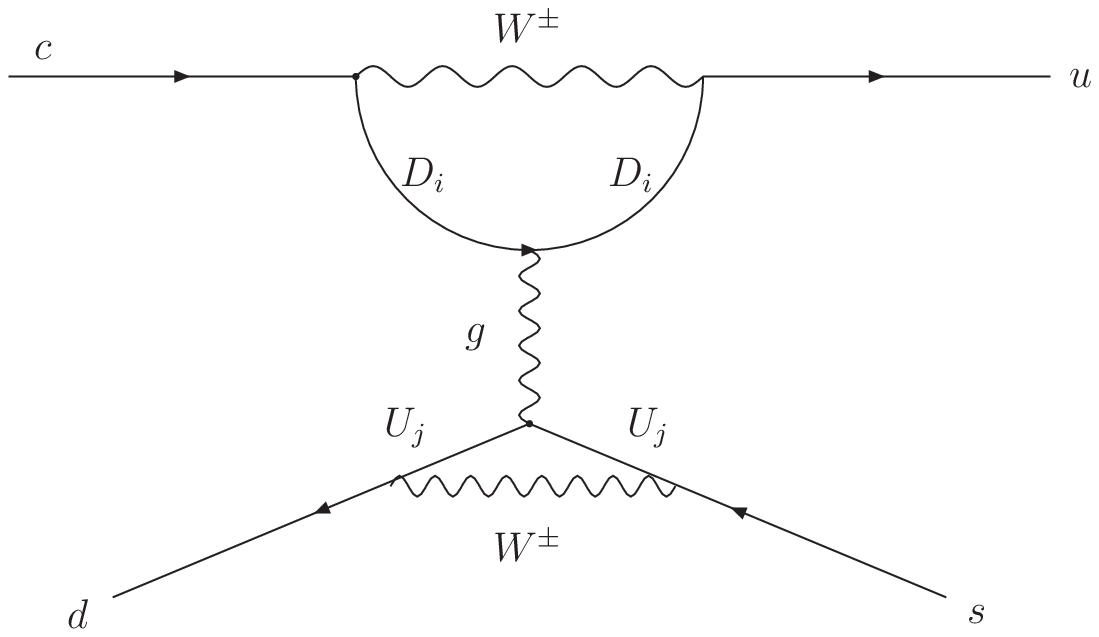}
\hspace{0.2cm}
\includegraphics[width=6cm,height=4cm]{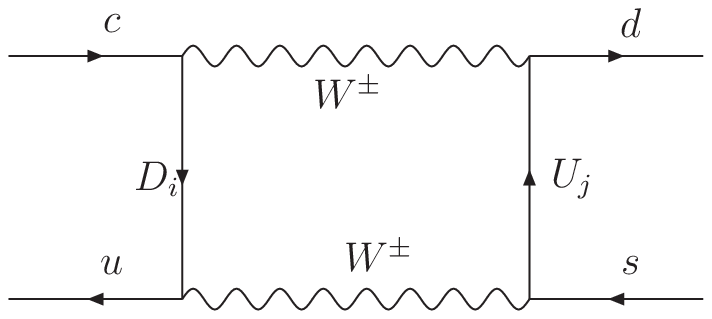}
\medskip
\caption{Feynman diagrams for DCS processes: left (right)
di-penguins contribution (box) contribution. For CF processes we
make the replacements $d \leftrightarrow s$ in each
diagram.\label{fig}}
\end{figure}

Upon evaluating the matrix elements of the operators in
Eq.(\ref{SMH}), we obtain the amplitude of the $\rm CF$ decay mode
$D^+ \rightarrow \bar K^0 \pi^+$ and $\rm DCS$ decay mode $D^+
\rightarrow K^0 \pi^+$ as

\begin{eqnarray}
A^{SM}_{D^+ \rightarrow \bar K^0 \pi^+} &=& -i{G_F\over
\sqrt{2}}V_{cs}^*V_{ud} \left[(a_1+\Delta a^{s d}_1)
X^{\pi^+}_{D^+ \bar K^0} + (a_2+\Delta a^{s d}_2+\Delta a^{s d
K^0}_2) X^{\bar K^0}_{D^+ \pi^+} \right],\nonumber\\
A^{SM}_{D^+ \rightarrow K^0 \pi^+} &=& i{G_F\over
\sqrt{2}}V_{cd}^*V_{us} \left[(a_1+\Delta a^{d s}_1)
X^{D^+}_{K^0\pi^+} + (a_2+\Delta a^{d s}_2-\Delta a^{d s K^0}_2)
X^{K^0}_{D^+ \pi^+} \right]\label{am00}
\end{eqnarray}

with  $ X^{P_1}_{P_2P_3}$ is given by
\begin{eqnarray}
X^{P_1}_{P_2P_3}= if_{P_1}\Delta_{P_2P_3}^2
F_0^{P_2P_3}(m_{P_1}^2),\  \Delta_{P_2P_3}^2=m_{P_2}^2-m_{P_3}^2
\end{eqnarray}
here $f_{P}$ is the $P$ meson  decay constant and $F_0^{P_2 P_3}$
is the form factor. In Eq.(\ref{am00}) the coefficients $a_1 =
c_1+ c_2/N_C $ and $a_2=-( c_2+c_1/N_C)$, where $N_C$ is the color
number, account for the tree-level contributions to the
amplitudes. These contributions originate from integrating out the
$W^\pm$ boson mediating the tree-level diagrams. On the other
hand, and in the same equations, $\Delta a^{s d, d s}_{1,2}$ and
$\Delta a^{s d K^0, d s K^0}_2 $ account for the contributions to
the amplitudes originating from integrating out the $W^\pm$ boson
mediating the box and di-penguin diagrams in Fig.\ref{fig}. Their
expressions can be obtained from the following expressions upon
setting $q =s $ and $q' = d$ for $\rm CF$ decay mode $D^+
\rightarrow \bar K^0 \pi^+$  and  $q =d $ and $q' = s$ for $\rm
DCS$ decay mode $D^+ \rightarrow K^0 \pi^+$

\begin{eqnarray}
\Delta a^{q q'}_1 &\simeq& -{G_Fm_W^2\over \sqrt{2}\ \pi^2V_{c
q}^*V_{u q'}N } {\cal B}^{q q'}_x- {G_F \alpha_S\over 4\sqrt{2}
 \pi^3V_{c q} ^* V_{u q'}}\left[{\kappa\over 2}\left(1-{1\over N^2}\right)\right]
 {\cal P}^{q q'}_g \nonumber \\
\Delta a^{q q'}_2 &\simeq& -{G_Fm_W^2\over \sqrt{2}\ \pi^2V_{c
q}^*V_{u q'} } {\cal B}^{q q'}_x\nonumber\\
\Delta a^{q q'K^0}_2 &\simeq& - {G_F \alpha_S\over 4\sqrt{2}
\pi^3V_{c q} ^* V_{u q'}}{3  m_{q'} m_c\over 8 N }\chi^{K^0} {\cal
P}^{q q'}_g
\end{eqnarray}

where $\kappa =(m_D^2+m_K^2)/2+3m_\pi^2/4$ and $\chi^{K^0}
=m_{K^0}^2/\big[(m_c-m_u)(m_d+m_s)\big]\simeq 2$. The quantities
${\cal B}^{q q'}_x $ and ${\cal P}^{q q'}_g$ originate from the
box and di-penguin diagrams respectively and they are given as
\bea {\cal B}^{q q'}_x &=& V_{c D}^*V_{u D} V_{U q}^*V_{U q'}
f(x_U,\
x_D)\nonumber\\
{\cal P}^{q q'}_g &=&\left[  V_{c D}^*V_{u D}  E_0(x_D)\right]
\left[ V_{U q}^*V_{U q'} E_0(x_U)\right]\eea with $U=u,\ c,\ t$
and $D=d,\ s,\ b$, $x_q=(m_q/m_W)^2$ and $f_{UD} \equiv
f(x_U,x_D)$ where \cite{inami}

\begin{eqnarray}
f(x,\ y) ={7xy-4\over 4(1-x)(1-y)} +{1\over x-y}\left[ {y^2\log
y\over (1-y)^2}\left(1-2x+{xy\over 4}\right)-  {x^2\log x\over
(1-x)^2}\left(1-2y+{xy\over 4}\right)   \right] \nonumber
\end{eqnarray}
and the Inami function $E_0(x)$ is given as
\begin{eqnarray}
E_0(x) &=& {1\over 12(1-x)^4}\left[
x(1-x)(18-11x-x^2)-2(4-16x+9x^2)\log(x)\right]
\end{eqnarray}

In NFA, there is no source for the strong CP conserving phases
required for having non vanishing direct CP aymmetries.
Consequently this factorization approximation is irrelevant to the
study of CP violation. On the other hand the mass of the charm
quark is not heavy enough to allow for a sensible heavy quark
expansion, such as in QCD factorization and soft collinear
effective theory, and it is not light enough for the application
of chiral perturbation theory \cite{Cheng:2010ry}.  A possible
approach to study charm decays in a model-independent way is the
so called the diagrammatic approach \cite{Chau:1982da,
Chau:1986du,Chau:1987tk,Chau:1989tk,Buccella:1994nf,
Cheng:2010ry}. Within this approach, the amplitude is decomposed
into  parts corresponding to generic quark diagrams according to
the topologies of weak interactions. For each one of these
topological diagrams, the related magnitude and relative strong
phase can be extracted from the data without making further
assumptions, apart from flavor SU(3) symmetry \cite{Cheng:2010ry}.

 In the diagrammatic approach the amplitudes of
 the the $\rm CF$ decay mode
$D^+ \rightarrow \bar K^0 \pi^+$ and  $\rm DCS$ decay mode $D^+
\rightarrow K^0 \pi^+$ can be written as \cite{Cheng:2010ry} \bea
A_{D^+ \rightarrow \bar K^0 \pi^+} &= &
V_{cs}^*V_{ud}(T+C)\nonumber\\
{\mathcal A}_{D^+ \rightarrow K^0 \pi^+} &=&
V^*_{cd}V_{us}(C''+A'') \label{dia}\eea

where $T$  represents the tree level color-allowed external
W-emission quark diagram, C and $C''$ denote the color-suppressed
internal W-emission diagram and $A''$ is the W-annihilation
diagram \cite{Cheng:2010ry}. Comparing Eq.(\ref{dia}) and
Eq.(\ref{am00}) we find that
\begin{eqnarray}
T& = & {G_F\over \sqrt{2}} (a_1+\Delta a^{s d}_1) f_\pi(m_D^2-m_K^2)F_0^{DK}(m_\pi^2) \\
C & = & {G_F\over \sqrt{2}} (a_2+\Delta a^{s d}_2+\Delta a^{s d
K^0}_2) f_K (m_D^2-m_\pi^2)F_0^{D\pi}(m_K^2) \nonumber\\ C''&=&
\frac{G_F}{\sqrt{2}} (a_1+\Delta a^{d s}_1)
f_{D}(m^2_K-m^2_\pi)F^{K\pi}_0(m^2_{D})\nonumber\\
E''&=& \frac{G_F}{\sqrt{2}} (a_2+\Delta a^{d s}_2-\Delta a^{d s
K^0}_2) f_K (m^2_D-m^2_{\pi})F^{D\pi}_0(m^2_K)\eea

 The direct CP asymmetry of the CF decay mode $D^+ \rightarrow \bar K^0 \pi^+$ can be expressed as

\begin{eqnarray}
A^{SM}_{CP} (D^+ \rightarrow \bar K^0 \pi^+) &=& {|A^{SM}_{D^+
\rightarrow \bar K^0 \pi^+}|^2-|\bar A^{SM}_{D^+ \rightarrow \bar
K^0 \pi^+}|^2 \over |A^{SM}_{D^+ \rightarrow \bar K^0
\pi^+}|^2+|\bar A^{SM}_{D^+ \rightarrow \bar K^0 \pi^+}|^2 }=
\kappa\, \sin(\phi_2-\phi_1)\label{acp1}
\end{eqnarray}

with \be \kappa = {2 r\sin(\alpha)\over |1+ r|^2 }\ee

here $r=|C/T|$ and $\alpha= \alpha_{C}- \alpha_{T}$.  The phases
$\alpha_{C}$ and $\alpha_{T}$ are the strong phase of the
amplitudes $C$ and $T$ respectively. The weak phases $\phi_1$ and
$\phi_2$ are defined through \bea \phi_1 &=& \tan^{-1}\big(
\frac{|\Delta a^{s d}_1|\sin\Delta \phi_1}{a_1+|\Delta a^{s
d}_1|\cos\Delta \phi_1}\big) \nonumber\\  \phi_2 &=&
\tan^{-1}\big( \frac{|\Delta a^{s d}_2+\Delta a^{s d
K^0}_2|\sin\Delta \phi_2}{a_2+|\Delta a^{s d}_2+\Delta a^{s d
K^0}_2|\cos\Delta \phi_2}\big)\label{CFphase}\eea

where the phases $\Delta \phi_1$ and $\Delta \phi_2$  are the
phase of $\Delta a^{s d}_1$ and $\Delta a^{s d}_2+\Delta a^{s d
K^0}_2$ respectively. Regarding the $\rm DCS$ decay mode $D^+
\rightarrow K^0 \pi^+$, the direct CP asymmetry can be expressed
as

\begin{eqnarray}
A^{SM}_{CP} (D^+ \rightarrow  K^0 \pi^+) &=& {|A^{SM}_{D^+
\rightarrow  K^0 \pi^+}|^2-|\bar A^{SM}_{D^+ \rightarrow  K^0
\pi^+}|^2 \over |A^{SM}_{D^+ \rightarrow  K^0 \pi^+}|^2+|\bar
A^{SM}_{D^+ \rightarrow  K^0 \pi^+}|^2 }= \kappa'\,
\sin(\phi'_2-\phi'_1)\label{acp2}
\end{eqnarray}

where  \be \kappa' = {2 r'\sin(\alpha')\over |1+ r'|^2 }\ee

with $r'=|A''/C''|$ and $\alpha'= \alpha_{A''}- \alpha_{C''}$. The
phases $\alpha_{A''}$ and $\alpha_{C''}$ are the strong phase of
the amplitudes $A''$ and $C''$ respectively. The weak phases
$\phi'_1$ and $\phi'_2$ are defined through \bea \phi'_1 &=&
\tan^{-1}\big( \frac{|\Delta a^{ d s}_1|\sin\Delta
\phi'_1}{a_1+|\Delta a^{d s}_1|\cos\Delta \phi'_1}\big) \nonumber\\
\phi_2 &=& \tan^{-1}\big( \frac{|\Delta a^{d s }_2-\Delta a^{d s
K^0}_2|\sin\Delta \phi'_2}{a_2+|\Delta a^{d s}_2-\Delta a^{d s
K^0}_2|\cos\Delta \phi'_2}\big)\label{DCSphase}\eea

here the phases $\Delta \phi'_1$ and $\Delta \phi'_2$  are the
phase of $\Delta a^{d s}_1$ and $\Delta a^{d s}_2+\Delta a^{d s
K^0}_2$ respectively.

Using he fitted values $T =  (3.14\pm 0.06)\cdot 10^{-6}{\rm
GeV}$, $C= C'' = \big(2.61\pm 0.08\big) \cdot 10^{-6}\cdot {\rm
e}^{-i (152\pm 1)^\circ}\ {\rm GeV}$ and $ A'' =
(0.39^{+0.13}_{-0.09})\times 10^{-6} e^{i(31^{+20}_{-33})\circ}$
\cite{Cheng:2010ry} we find that $\kappa \simeq -0.23$ and
$\kappa' \simeq -0.01$. On the other hand using $a_1= 1.2 \pm 0.1$
and $a_2 = - 0.5 \pm 0.1$ and the expressions in
Eqs.(\ref{CFphase},\ref{DCSphase}) we find that
 $\sin(\phi_2-\phi_1) \simeq -2.0\times 10^{-10}$ and
$\sin(\phi'_2-\phi'_1) \simeq 4.8\times 10^{-9}$. Thus, from
Eqs.(\ref{acp1}, \ref{acp2}) we find that $A^{SM}_{CP} (D^+
\rightarrow \bar K^0 \pi^+)\simeq 4.6 \times 10^{-11} $ and
$A^{SM}_{CP} (D^+ \rightarrow  K^0 \pi^+)\simeq -5.7 \times
10^{-11} $. Clearly the predicted direct CP asymmetries within SM
are so tiny in both decay modes due to the highly suppressed
generated weak phases originating  leaving a room for a possible
enhancement from new physics beyond SM that have new weak phases.

\section{Models with Charged Higgs contributions}\label{Hig}

Possible extensions of the SM include the two Higgs doublet models
(2HDM)\cite{Haber:1978jt,Abbott:1979dt}. Based on their couplings
to quarks and leptons, these models can be classified to several
types such as: type I, II or III (for a review see ref.
\cite{Branco:2011iw}). Among these types 2HDM type III (2HDM III)
is of a particular interest to our study. This can be attributed
to the presence of complex couplings of Higgs to quarks. These
couplings are relevant for generating the desired CP violating
weak phases. In the literature,  2HDM III has gain interest as it
can explain $B\to D\tau\nu$, $B\to D^*\tau\nu$ and $B\to \tau\nu$
simultaneously while other types such as 2HDM I and 2HDM II cannot
\cite{Crivellin:2012ye}.

In  2HDM III the physical mass eigenstates are $H_0$ (heavy
CP-even Higgs), $h_0$ (light CP-even Higgs) and $A_0$ (CP-odd
Higgs) and $H^{\pm}$. In this model both Higgs doublets can couple
to up-type and down-type  quarks. As a consequence the couplings
of the neutral Higgs mass eigenstates can induce flavor violation
in Neutral Currents at tree-level. In the down sector these flavor
violating couplings are stringently constrained from flavor
changing neutral current processes
\cite{Crivellin:2012ye,Crivellin:2013wna}. Thus in the following
we consider only charged Higgs couplings to quarks that can be
expressed as \cite{Crivellin:2010er,Crivellin:2012ye}:
\begin{equation}
\mathcal{L}^{eff}_{H^\pm} = \bar{u}_f {\Gamma_{u_f d_i
}^{H^\pm\,LR\,\rm{eff} } }P_R d_i
+ \bar{u}_f {\Gamma_{u_f d_i }^{H^\pm\,RL\,\rm{eff} } }P_L d_i\, ,\\
 \label{Higgs-vertex}
\end{equation}
where \bea {\Gamma_{u_f d_i }^{H^\pm\,LR\,\rm{eff} } } &=&
\sum\limits_{j = 1}^3 {\sin\beta\, V_{fj} \left( \frac{m_{d_i
}}{v_d} \delta_{ji}-
  \epsilon^{ d}_{ji}\tan\beta \right), }
\nonumber\\
{\Gamma_{u_f d_i }^{H^ \pm\,RL\,\rm{eff} } } &=& \sum\limits_{j =
1}^3 {\cos\beta\,  \left( \frac{m_{u_f }}{v_u} \delta_{jf}-
  \epsilon^{ u\star}_{jf}\tan\beta \right)V_{ji}}
 \label{Higgsv}
\eea Here $v_u$ and $v_d$ denote the vacuum expectations values of
the neutral component of the  Higgs doublets,  tan $\beta =
v_u/v_d$ and $V$ is the CKM matrix. Applying the Feynman-rules
given in Eq.(\ref{Higgs-vertex}) allows us to calculate the
contributions to the total amplitude originating from tree-level
Charged Higgs mediation.

The contribution of charged Higgs to the effective Hamiltonian can
be written as

\bea
Q^{H^\pm}_1 &=&(\bar{q} P_R c)(\bar{u} P_L q'),\nonumber\\
Q^{H^\pm}_2 &=&(\bar{q} P_L c)(\bar{u} P_R q'),\nonumber\\
Q^{H^\pm}_3 &=&(\bar{q} P_L c)(\bar{u} P_L q'),\nonumber\\
Q^{H^\pm}_4 &=&(\bar{q} P_R c)(\bar{u} P_R q'),
 \eea

where as before for $\rm CF$ decays  $q =s $ and  $q' = d$ while
for $\rm DCS$ decays we have $q =d $ and  $q' = s$. The Wilson
coefficients $C^H_i$, at the electroweak scale, can be expressed
as

\begin{eqnarray}
C^{H^\pm}_1 &=& \frac {\sqrt{2} }{ G_F V^*_{c q}V_{u q'}
 m^2_H} \bigg(\sum\limits_{j = 1}^3
{\cos\beta\, V_{j q'} \left( \frac{m_u }{v_u} \delta_{j1}-
\epsilon^{ u\star}_{j1}\tan\beta \right)}\bigg)\bigg(
\sum\limits_{k= 1}^3 {\cos\beta\,V^{\star}_{k q}} \left(
\frac{m_c}{v_u} \delta_{k2}-\epsilon^{ u}_{k2}\tan\beta
\right)\bigg),\nonumber\\
C^{H^\pm}_2 &=& \frac {\sqrt{2} }{ G_F V^*_{c q}V_{u q'}
 m^2_H} \bigg(\sum\limits_{j = 1}^3 {\sin\beta\,V_{1j}  \left( \frac{m_{q'
}}{v_d} \delta_{j q'} - \epsilon^{ d}_{j q'}\tan\beta
\right)}\bigg) \bigg( \sum\limits_{k= 1}^3
{\sin\beta\,V^{\star}_{2k}} \left( \frac{m_{q}}{v_d} \delta_{k
q}-\epsilon^{ d\star}_{k q}\tan\beta
\right)\bigg),\nonumber\\
 C^{H^\pm}_3 &=& \frac {\sqrt{2} }{  G_F V^*_{c q}V_{u q'}
 m^2_H} \bigg(\sum\limits_{j = 1}^3 {\cos\beta\, V_{j q'} \left(
\frac{m_u }{v_u} \delta_{j1}- \epsilon^{ u\star}_{j1}\tan\beta
\right)}\bigg) \bigg( \sum\limits_{k= 1}^3
{\sin\beta\,V^{\star}_{2k}} \left( \frac{m_{q}}{v_d} \delta_{k
q}-\epsilon^{ d\star}_{k q}\tan\beta
\right)\bigg),\nonumber\\
C^{H^\pm}_4 &=& \frac {\sqrt{2} }{ G_F V^*_{c q}V_{u q'}
 m^2_H}\bigg( \sum\limits_{k= 1}^3
{\cos\beta\,V^{\star}_{k q}} \left( \frac{m_c}{v_u}
\delta_{k2}-\epsilon^{ u}_{k2}\tan\beta \right)\bigg)
\bigg(\sum\limits_{j = 1}^3 {\sin\beta\,V_{1j}  \left( \frac{m_{q'
}}{v_d} \delta_{j q'} -
\epsilon^{ d}_{j q'}\tan\beta \right)}\bigg)\nonumber \\
 \label{Higgsw}
\end{eqnarray}

In order to evaluate the contributions of the charged Higgs to the
amplitudes of the decay modes under consideration we need to
discuss the restraints imposed on the flavor-changing parameters
$\epsilon^{u,d}_{ij}$ appear in the expressions of $C^{H^\pm}_i$
above. We consider first the down sector and discuss the possible
constraints that can be imposed on $\epsilon^{d}_{ij}$. For the
case $i\neq j$, stringent bounds on $\epsilon^{d}_{ij}$ from
considering flavor changing neutral current (FCNC) processes  due
to the tree level neutral Higgs exchange
\cite{Crivellin:2012ye,Crivellin:2013wna}. As a result, they
cannot contribute significantly to the decay modes under
investigation. Thus we are left with $\epsilon^{d}_{11}$,
$\epsilon^{d}_{22}$ and $\epsilon^{d}_{33}$. The couplings
$\epsilon^{d}_{11}$ and $\epsilon^{d}_{22}$ can be severely
constrained by applying the naturalness criterion of 't Hooft to
the quark masses. In view of the criterion, the smallness of a
quantity is only natural if a symmetry is gained in the limit in
which this quantity is zero \cite{Crivellin:2012ye}. Consequently,
it is unnatural to have large accidental cancellations without a
symmetry forcing these cancellations. Applying the naturalness
criterion to the quark masses leads to the bounds given as
\cite{Crivellin:2013wna}

\begin{eqnarray}
|v_{u(d)} \epsilon^{d(u)}_{ij}|&\leq&
\left|V^{CKM}_{ij}\right|\,{\rm max
}\left[m_{d_i(u_i)},m_{d_j(u_j)}\right]\,\,\,\,\,\,\,\, for\, i <
j\nonumber\\
|v_{u(d)} \epsilon^{d(u)}_{ij}|&\leq&{\rm max
}\left[m_{d_i(u_i)},m_{d_j(u_j)}\right]\, \,\,\,\,\,\,\,\,
\,\,\,\,\, \,\,\,\,\, \,\,\,\,\, \,\,\,\,\,  for\, i \geq
j.\label{constr}
\end{eqnarray}

Clearly, due to the smallness of the $d$ and $s$ quark masses, the
constraints on  $\epsilon^d_{11}$ and $\epsilon^d_{22}$ are so
strong. Thus we are left with $\epsilon^d_{33}$ which is
irrelevant to the decay modes we are interested in. Putting all
together, we can safely neglect terms proportional to the
couplings $\epsilon^{d}_{ij}$ in $C^{H^\pm}_i$.

We turn now to discuss the constraints that can be set on the
couplings $\epsilon^{u}_{ij}$. Again, applying the naturalness
criterion of 't Hooft to the $u$ quark mass we find that, using
second line of Eq.(\ref{constr}), the constraint on
$\epsilon^u_{11}$ is so severe.  As a result we can drop terms
proportional to $\epsilon^{u}_{11}$ in $C^{H^\pm}_i$. Thus, to a
good approximation, we can finally write
\begin{eqnarray}
C^{H^\pm}_1 &\simeq& -\frac {\sin 2\beta V_{3 q'} \epsilon^{
u\star}_{31}}{\sqrt{2} \, G_F V^*_{c q}V_{u q'} m^2_H} \bigg(
\frac{m_c}{v_u} V^{\star}_{2 q}-\epsilon^{ u}_{22}\tan\beta
V^{\star}_{2 q} -\epsilon^{ u}_{32}\tan\beta V^{\star}_{3 q}\bigg),\nonumber\\
C^{H^\pm}_4 &\simeq& \frac {\sin 2\beta V_{1q'} m_{q' } }{
\sqrt{2} \, G_F V^*_{c q}V_{u q'} m^2_H v_d}\bigg( \frac{m_c}{v_u}
V^{\star}_{2 q}-\epsilon^{ u}_{22}\tan\beta V^{\star}_{2 q}
-\epsilon^{
u}_{32}\tan\beta V^{\star}_{3 q}\bigg),\nonumber \\
C^{H^\pm}_2 &\simeq& C^{H^\pm}_3 \simeq 0 \label{Higgsw}
\end{eqnarray}

where we have neglected the terms that are proportional to
$\epsilon^{ u}_{12} \epsilon^{ u\star}_{21}$ due to the strong
constraint $\mid\epsilon^u_{12}\epsilon^{u\,*}_{21}\mid\ < 2\times
10^{-8}$ from $D-\bar D$ mixing \cite{Crivellin:2013wna}.
Moreover, the bound also implies that $\mid\epsilon^u_{12,21}\mid
< \sqrt{2}\times 10^{-4}$ in the absence of a symmetry that
protect one of these parameters from being much smaller than the
other one. As a consequence, we neglected terms proportional to
$\epsilon^u_{12}$ in the above Wilson coefficients. We also
neglected terms suppressed by the up quark mass.

We proceed now to calculate the amplitudes of the decay processes
of interest.  For $\rm CF$ decay modes $D^+ \rightarrow \bar K^0
\pi^+$, the total amplitude, including Higgs contribution, can be
written as
\begin{eqnarray}
A^{SM+H^\pm}_{D^+ \rightarrow \bar K^0 \pi^+} &\simeq& -i{G_F\over
\sqrt{2}}V_{cs}^*V_{ud} \left[(a_1+\Delta a^{H}_1) X^{\pi^+}_{D^+
\bar K^0} + (a_2+\Delta a^{H \,\bar K^0}_2) X^{\bar K^0}_{D^+
\pi^+} \right],\nonumber\\\label{am1111}
\end{eqnarray}
with \be \Delta a^{H}_1 = \chi^{\pi^+}(\mathcal C^H_1- \mathcal
C^H_4),\,\,\,\,\,\,\,\,\,\,\,\,\,\,\,\,\,\,\,\,\,\,\,\, \Delta
a^{H \,\bar K^0}_2 = \frac{1}{2N}\big(\mathcal C^H_1-\chi^{ K^0}
\mathcal C^H_4\big)\ee The quantities $\mathcal C^H_{1.4}$ can be
obtained from $C^H_{1,4}$ by setting $q=s$ and $q'=d$ and \bea
\chi^{\pi^+}&=&{m_{\pi}^2\over (m_c-m_s)(m_u+m_d)}\eea

In the case of $\rm DCS$ decay modes $D^+ \rightarrow K^0 \pi^+$,
the total amplitude can be expressed as
\begin{eqnarray}
A^{SM+H^\pm}_{D^+ \rightarrow K^0 \pi^+} &=& i{G_F\over
\sqrt{2}}V_{cd}^*V_{us} \left[(a_1+\Delta a^{H\,D^+}_1)
X^{D^+}_{K^0\pi^+} + (a_2+\Delta a^{H\,K^0}_2) X^{K^0}_{D^+ \pi^+}
\right],\nonumber\\\label{am00000}
\end{eqnarray}
where \bea \Delta a^{H\,D^+}_1 &=& \chi^{D^+}(\mathcal C'^H_1+
\mathcal C'^H_4),\,\,\,\,\,\,\,\,\,\,\,\,\,\,\,\,\,\,\,\,\,\,\,\,
\Delta a^{H\,K^0}_2 = \frac{1}{2N}\big(\mathcal C'^H_1-\chi^{ K^0}
\mathcal C'^H_4\big)\eea

The quantities $\mathcal C'^H_{1.4}$ can be obtained from
$C^H_{1,4}$ by setting $q=d$ and $q'=s$ and

\bea \chi^{D^+} = {m_{D^+}^2\over (m_c+m_d)(m_u-m_s)},\eea

In a recent study a lower bound on the charged Higgs mass in 2HDM
of Type II has been set after taking into account all relevant
results from direct charged and neutral Higgs boson searches at
LEP and the LHC,  as well as  the most recent constraints from
flavour physics \cite{Arbey:2017gmh}. The bound reads $
m_{H^\pm}\gtrsim 600$ GeV independent of $\tan \beta$. This bound
should be also respected in 2HDM III \cite{Crivellin:2012ye}.

For $\tan\beta = 50$, $m_H = 600$ GeV and keeping only dominant
terms, after considering constraints imposed on the $\epsilon^{
q}_{i j}$ studied in details in Ref.\cite{Crivellin:2013wna},  we
find that \bea \Delta
a^{H}_1&\simeq&  0.001 \, \epsilon^{ u}_{22} \nonumber\\
\Delta a^{H \,\bar K^0}_2 &\simeq&  0.0001\, \epsilon^{ u}_{22}\nonumber\\
\Delta a^{H\,D^+}_1&\simeq&   0.278 \, \epsilon^{ u}_{22}
\nonumber\\
\Delta a^{H \, K^0}_2 &\simeq& 0.003\, \epsilon^{ u}_{22} \eea

We proceed now to discuss the constraints imposed on the coupling
$\,\epsilon^{ u}_{22}$. The processes $D_{(s)} \to \tau \nu $,
$D_{(s)} \to \mu \nu $ can constraint the real part of
$\,\epsilon^{ u}_{22}$ while the constraints on the imaginary part
of $\,\epsilon^{ u}_{22}$ are weak \cite{Crivellin:2013wna}.
Regarding the imaginary part of $\,\epsilon^{ u}_{22}$ which is
relevant for generating direct CP violation, and for $m_{H^\pm} =
600$ GeV, $\tan\beta=50$, the constraints from the electric dipole
moment of the neutron reads $ -0.16 \lesssim \, Im(\epsilon^{
u}_{22})\, \lesssim 0.16$ \cite{Crivellin:2013wna}. Other
processes such as $D-\bar D$ mixing and $K-\bar K$ mixing can be
used to set bounds on $\,\epsilon^{ u}_{22}$. However these bounds
are weaker than the bounds obtained from  $D_{(s)} \to \tau \nu $,
$D_{(s)} \to \mu \nu $ and the electric dipole moment of the
neutron \cite{Delepine:2012xw,Crivellin:2013wna}.

 The real parts of $\Delta a^{H}_1$ and $\Delta a^{H}_2$ are
expected to be much smaller than the SM contributions, $a_1$ and
$a_2$, and hence we can be safely neglect them and keep only the
imaginary parts required for generating the weak phases.


 The direct CP asymmetry of the CF decay mode $D^+ \rightarrow \bar K^0 \pi^+$
, including Higgs contributions,  can be expressed as

\begin{eqnarray}
A^{SM+H}_{CP} (D^+ \rightarrow \bar K^0 \pi^+) &=&
{|A^{^{SM+H}}_{D^+ \rightarrow \bar K^0 \pi^+}|^2-|\bar
A^{^{SM+H}}_{D^+ \rightarrow \bar K^0 \pi^+}|^2 \over
|A^{^{SM+H}}_{D^+ \rightarrow \bar K^0 \pi^+}|^2+|\bar
A^{^{SM+H}}_{D^+ \rightarrow \bar K^0 \pi^+}|^2 }= \kappa\,
\sin(\phi^{H}_2-\phi^{H}_1)\label{acph1}
\end{eqnarray}

where where $\kappa$ is given as before and the weak phases
$\phi^{H}_1$ and $\phi^{H}_2$ are defined through \bea \phi^{H}_1
&=& \tan^{-1}\big(\frac{|\Delta a^H_1|\sin\Delta
\phi^H_1}{a_1}\big) \nonumber\\\phi^{H}_2 &=&
\tan^{-1}\big(\frac{|\Delta a^{H \,\bar K^0}_2|\sin\Delta
\phi^H_2}{a_2}\big)\eea   where $\Delta \phi^{H}_1$  and $\Delta
\phi^{H}_2$ are the phases of $\Delta a^{H}_1$ and $\Delta a^{H
\,\bar K^0}_2$ respectively. Regarding the $\rm DCS$ decay mode
$D^+ \rightarrow K^0 \pi^+$, the direct CP asymmetry can be
expressed as

\begin{eqnarray}
A^{SM+H}_{CP} (D^+ \rightarrow  K^0 \pi^+) &=& {|A^{^{SM+H}}_{D^+
\rightarrow  K^0 \pi^+}|^2-|\bar A^{^{SM+H}}_{D^+ \rightarrow K^0
\pi^+}|^2 \over |A^{^{SM+H}}_{D^+ \rightarrow  K^0 \pi^+}|^2+|\bar
A^{^{SM+H}}_{D^+ \rightarrow  K^0 \pi^+}|^2 }= \kappa'\,
\sin(\phi'^{H}_2-\phi'^{H}_1)\label{acph1}
\end{eqnarray}

where where $\kappa'$ is given as before and the weak phases
$\phi'^{H}_1$ and $\phi'^{H}_2$ are defined through \bea
\phi'^{H}_1 &=& \tan^{-1}\big(\frac{|\Delta a^{H D^+}_1|\sin\Delta
\phi'^H_1}{a_1}\big) \nonumber\\\phi'^{H}_2 &=&
\tan^{-1}\big(\frac{|\Delta a^{H \, K^0}_2|\sin\Delta
\phi'^H_2}{a_2}\big)\eea   where $\Delta \phi'^{H}_1$  and $\Delta
\phi'^{H}_2$ are the phases of $\Delta a^{H D^+}_1$ and $\Delta
a^{H \, K^0}_2$ respectively.  Assuming maximum value of
$Im(\epsilon^{ u}_{22})$, we obtain CP asymmetry $A^{SM+H}_{CP}
(D^+ \rightarrow \bar K^0 \pi^+)\simeq 3.8 \times 10^{-5} $ and
$A^{SM+H}_{CP} (D^+ \rightarrow  K^0 \pi^+)\simeq 4.5 \times
10^{-4} $. This result show that charged Higgs contributions to
the amplitudes of these decay modes can enhance the direct CP
asymmetry 6 and 7 orders of magnitudes for the CF decay mode $D^+
\rightarrow \bar K^0 \pi^+$ and the $\rm DCS$ decay mode $D^+
\rightarrow K^0 \pi^+$ respectively.



\section{A new charged gauge boson as Left Right models \label{LRSS}}

In this section, we consider a new physics model based on the
gauge group $SU(2)_L \times SU(2)_R \times U(1)_{B-L}$
\cite{Pati:1973rp,Mohapatra:1974hk,Mohapatra:1974gc,Senjanovic:1975rk,Senjanovic:1978ev,Beall:1981ze,Cocolicchio:1988ac,Langacker:1989xa,Cho:1993zb,Babu:1993hx}.
Assuming no mixing between $W_L$ and $W_R$ gauge bosons, the
contributions from new diagrams, similar to the SM tree-level
diagrams with $W_L $ is replaced by a $W_R$, to the effective
Hamiltonian governs  $D \to K \pi$ decays can be expressed as :
\begin{eqnarray}
{\cal H}_{\rm LR} &=& {G_F\over\sqrt{2}}\left({g_R m_W\over g_L
m_{W_R}}\right)^2V_{R c q}^*V_{R u q'}\bigg[c_1'\big(\bar q
\gamma_\mu c_R\big)  \big(\bar u\gamma^\mu q'_R\big) + c_2' (\bar
u \gamma_\mu c_R\bar q \gamma^\mu q'_R)\bigg]+{\rm h.c.}
\end{eqnarray}
here $g_{L}$ and $g_{R}$ denote the gauge $SU(2)_{L}$ and
$SU(2)_{R}$ couplings respectively and $q$ and $q'$ can be
different light down-type quarks. The masses $m_W$ and $m_{W_R}$
represent the $SU(2)_{L}$ and $SU(2)_{R}$ charged gauge boson
masses respectively and $ V_R$ is the quark mixing matrix in the
right sector in analogy to the CKM quark mixing matrix, $V_{CKM
}\equiv V$, in the left sector of the charged quark currents.
ATLAS and CMS have set stringent limits on $m_{ W_R}$, in the
$3.5-4.4 $ $ \rm TeV$ region based on their latest analyses with
$37\, \rm f b^{- 1}$ and $35.9\, \rm f b^{- 1}$ luminosities,
respectively, at $ \sqrt{s} = 13 $ $ \rm TeV$
\cite{Aaboud:2017yvp,Sirunyan:2017ukk,Sirunyan:2017vkm,Sirunyan:2018xlo,Sirunyan:2018pom,Aaboud:2018spl}.
These analyses rely on the assumptions that the model is
manifestly left-right symmetric i.e. $g_L = g_R$ and that $V_R$ is
either diagonal, or $V_R = V$. Clearly, due to the stringent
limits on $m_{W_R}$ and the assumptions of $V_R$, one expects that
no sizeable CP asymmetry can be obtained in this class of left
right symmetric models for both $\rm CF$ and $\rm DCS$ decay modes
of $D \to K \pi$ decays.

Previous studies showed that sizable CP asymmetries can be
obtained in the Charm and muon sectors in a general left right
symmetric model \cite{Chen:2012usa,Lee:2011kn,Delepine:2012xw}. In
this model, the mixing between the left and the right gauge bosons
is allowed and the left-right symmetry  is not manifest at
unification scale. In order to estimate the CP asymmetries of the
$\rm CF$ and $\rm DCS$ decay modes of $D \to K \pi$ decays, in the
framework of this model, we start by parameterizing the charged
current mixing matrix as \cite{Herczeg:1985cx,
Langacker:1989xa,Chen:2012usa}
\begin{eqnarray}
\left(\begin{array}{c}
  W^\pm_L \\
  W^\pm_R
\end{array}\right) =
\left(\begin{array}{cc}
  \cos \xi & -\sin \xi \\
{\rm e}^{i\omega}\sin \xi & {\rm e}^{i\omega}\cos \xi
\end{array}\right)
\left(\begin{array}{c}
  W^\pm_1 \\
  W^\pm_2
\end{array}\right)\simeq
 \left(\begin{array}{cc}
  1 & - \xi \\
{\rm e}^{i\omega}\xi & {\rm e}^{i\omega}
\end{array}\right)
\left(\begin{array}{c}
  W^\pm_1 \\
  W^\pm_2
\end{array}\right)
\end{eqnarray}
where $\xi$ is a mixing angle,  $W^\pm_1$ and $W^\pm_2$ are the
mass eigenstates and $\omega$ is a CP violating phase. Hence, the
charged currents interaction in the quark sector can be expressed
as
\begin{eqnarray}
{\cal L} &\simeq & -{1\over \sqrt{2}} \bar U \gamma_\mu
\left(g_LVP_L+g_R\xi \bar V^RP_R\right)DW_1^\dagger- {1\over
\sqrt{2}} \bar U \gamma_\mu \left(-g_L\xi VP_L+g_R\bar
V^RP_R\right)DW_2^\dagger
\end{eqnarray}
where $\bar V^R={\rm e}^{i\omega}V^R$. Upon integrating out $W_1$
in the usual way and neglecting $W_2$ contributions, given its
mass is much higher, we obtain the effective hamiltonian relevant
to the $\rm CF$ and $\rm DCS$ $D \to K \pi$ decays as:
\begin{eqnarray}
{\cal H}^{q q'}_{\rm eff.} &=& {4G_F\over \sqrt{2}}\bigg\{c_1
\bigg[\ \bar{q}\gamma_\mu \big(V^*P_L+{g_R\over g_L}\xi \bar
V^{R*}P_R \big)_{c\, q} c \bigg] \bigg[ \bar{u}\gamma^\mu
\big(VP_L+{g_R\over g_L}\xi \bar VP_R  \big)_{u \, q'}  q'\bigg]
\nonumber  \\
&+&  c_2 \bigg[ \bar q_\alpha \gamma_\mu \big(V^*P_L+{g_R\over
g_L}\xi \bar V^{R*}P_R  \big)_{c\, q} c_\beta\bigg] \bigg[\bar
u_\beta\gamma^\mu \big(VP_L+{g_R\over g_L}\xi \bar VP_R \big)_{u\,
q'} q'_\alpha\bigg] \bigg\}+{\rm h.\ c.}
\nonumber \\
\end{eqnarray}
where $\alpha,\beta$ are color indices and $q,q'$ are different
light down-type quarks. The terms of the effective Hamiltonian
proportional to $\xi$ are: \bea \Delta {\cal H}^{q q'}_{\rm
eff}&\simeq & {G_F\over \sqrt{2}}{g_R\over g_L}\xi \left[c_1\bar q
\gamma_\mu V_{c \, q}^*c_L\bar u \gamma^\mu \bar V^R_{u\,q'} q'_R+
c_1 \bar q \gamma_\mu \bar V^{R*}_{c\,q} c_R \bar u \gamma^\mu
V_{u\,q'} q'_L \right.
\nonumber  \\
&+& \left. c_2\bar q_\alpha \gamma_\mu V_{c\,q }^*c_{L\beta} \bar
u_\beta \gamma^\mu \bar V^R_{u\,q '} q'_{R\alpha}+ c_2 \bar
q_\alpha \gamma_\mu \bar V^{R*}_{c\,q } c_{R\beta} \bar u_\beta
\gamma^\mu V_{u\,q'} q'_{L\alpha} \right]+{\rm h.\ c.}
\label{LRS}\eea Upon evaluating the matrix elements of the
operators in Eq.(\ref{LRS}), we  obtain the new contribution to
the amplitude of the $\rm CF$ decay mode $D^+ \rightarrow \bar K^0
\pi^+$ by setting $q=s$ and $q'=d$
\begin{eqnarray}
A^{LR}_{D^+ \rightarrow \bar K^0 \pi^+} &= & -{iG_F\over
\sqrt{2}}{g_R\over g_L}\xi \left[-c_1 V_{cs}^*\bar V^R_{ud}
 \left( X^{\pi^+}_{D^+ \bar K^0} -{2\over N}\chi^{K^0} X^{\bar K^0}_{D^+ \pi^+}  \right)+ c_1 \bar V^{R*}_{cs} V_{ud}
 \left( X^{\pi^+}_{D^+ \bar K^0} - {2\over N}\chi^{K^0} X^{\bar K^0}_{D^+ \pi^+}  \right) \right.
\nonumber  \\
&& \left. -c_2 V_{cs}^*\bar V^R_{ud} \left(- 2\chi^{K^0} X^{\bar
K^0}_{D^+ \pi^+} +{1\over N} X^{\pi^+}_{D^+ \bar K^0}  \right)+
c_2\bar V^{R*}_{cs}V_{ud} \left(-2 \chi^{K^0} X^{\bar K^0}_{D^+
\pi^+} +{1\over N} X^{\pi^+}_{D^+ \bar K^0} \right)\right]
\nonumber  \\
&= & {iG_F\over \sqrt{2}}{g_R\over g_L}\xi \left(V_{cs}^*\bar
V^R_{ud}-\bar V^{R*}_{cs} V_{ud}\right) \left( a_1 X^{\pi^+}_{D^+
\bar K^0} - 2\chi^{K^0}  a_2 X^{\bar K^0}_{D^+ \pi^+}  \right)
\end{eqnarray}
and thus, the total amplitude, including SM contribution, can be
written as
\begin{eqnarray}
A^{SM+LR}_{D^+ \rightarrow \bar K^0 \pi^+} &\simeq& -i{G_F\over
\sqrt{2}}V_{cs}^*V_{ud} \left[(a_1+\Delta a^{LR}_1) X^{\pi^+}_{D^+
\bar K^0} + (a_2+\Delta a^{LR \,\bar K^0}_2) X^{\bar K^0}_{D^+
\pi^+} \right],\label{amLR1}
\end{eqnarray}
with \be \Delta a^{LR}_1 \simeq - {g_R\over g_L }\xi \left(\bar
V^R_{ud}-\bar V^{R*}_{cs} \right) a_1
,\,\,\,\,\,\,\,\,\,\,\,\,\,\,\,\,\,\,\,\,\,\,\,\, \Delta a^{LR
\,\bar K^0}_2 \simeq  {2 g_R\over g_L }\xi \left(\bar
V^R_{ud}-\bar V^{R*}_{cs} \right) \chi^{K^0} a_2 \label{del1}\ee
The direct CP asymmetry of the CF decay mode $D^+ \rightarrow \bar
K^0 \pi^+$ , including the new contributions,  can be expressed as
\begin{eqnarray}
A^{SM+LR}_{CP} (D^+ \rightarrow \bar K^0 \pi^+) &=&
{|A^{^{SM+LR}}_{D^+ \rightarrow \bar K^0 \pi^+}|^2-|\bar
A^{^{SM+LR}}_{D^+ \rightarrow \bar K^0 \pi^+}|^2 \over
|A^{^{SM+LR}}_{D^+ \rightarrow \bar K^0 \pi^+}|^2+|\bar
A^{^{SM+LR}}_{D^+ \rightarrow \bar K^0 \pi^+}|^2 }= \kappa\,
\sin(\phi^{LR}_2-\phi^{LR}_1)\label{acpLR1}
\end{eqnarray}
where where $\kappa$ is given as before and the weak phases
$\phi^{LR}_1$ and $\phi^{LR}_2$ are defined through \bea
\phi^{LR}_1 &=& \tan^{-1}\big(\frac{|\Delta a^{LR}_1|\sin\Delta
\phi^{LR}_1}{a_1}\big) \nonumber\\\phi^{LR}_2 &=&
\tan^{-1}\big(\frac{|\Delta a^{LR \,\bar K^0}_2|\sin\Delta
\phi^{LR}_2}{a_2}\big)\eea   where $\Delta \phi^{LR}_1$  and
$\Delta \phi^{LR}_2$ are the phases of $\Delta a^{LR}_1$ and
$\Delta a^{LR \,\bar K^0}_2$ respectively. We turn now to the $\rm
DCS$ decay mode $D^+ \rightarrow K^0 \pi^+$. proceeding in a
similar way as before, upon evaluating the matrix elements of the
operators in Eq.(\ref{LRS}) and setting $q=d$ and $q'=s$ we find
that the new contribution to the amplitude can be given as

\begin{eqnarray}
 A^{SM+LR}_{D^+ \rightarrow K^0 \pi^+} &= & -{iG_F\over \sqrt{2}}{g_R\over g_L}\xi \left[- c_1 V_{c d}^*\bar V^R_{u s}
 \left( X^{D^+}_{K^0\pi^+}-{2\over N}\chi^{K^0} X^{K^0}_{D^+\pi^+} \right)+ c_1
 \bar V^{R*}_{c d} V_{u s}  \left( X^{D^+}_{K^0\pi^+}-{2\over N}\chi^{K^0} X^{K^0}_{D^+\pi^+} \right) \right.
\nonumber  \\
&& \left. -c_2 V_{c d}^*\bar V^R_{u s} \left( - 2\chi^{K^0}
X^{DK^0}_{D^+\pi^+}+{1\over N} X^{D^+}_{K^0\pi^+} \right)+ c_2\bar
V^{R*}_{c d}V_{u s} \left(- 2 \chi^{K^0}
X^{K^0}_{D^+\pi^+}+{1\over N} X^{D^+}_{K^0\pi^+}\right)\right]
\nonumber  \\
&= & {iG_F\over \sqrt{2}}{g_R\over g_L}\xi \left(V_{c d}^*\bar
V^R_{u s}-\bar V^{R*}_{c d} V_{u s}\right)\left( a_1
X^{D^+}_{K^0\pi^+}- 2\chi^{K^0}  a_2 X^{K^0}_{D^+\pi^+} \right)
\end{eqnarray}
Thus, the total amplitude after including SM contribution can be
expressed as
\begin{eqnarray}
A^{SM+LR}_{D^+ \rightarrow K^0 \pi^+} &=& i{G_F\over
\sqrt{2}}V_{cd}^*V_{us} \left[(a_1+\Delta a'^{LR}_1)
X^{D^+}_{K^0\pi^+} + (a_2+\Delta a'^{LR\,K^0}_2) X^{K^0}_{D^+
\pi^+} \right],\label{amLR10}
\end{eqnarray}
where \bea \Delta a'^{LR}_1 &\simeq& {g_R\over g_L \lambda}\xi
\left(\bar V^R_{u s}+\bar V^{R*}_{c d} \right) a_1
,\,\,\,\,\,\,\,\,\,\,\,\,\,\,\,\,\,\, \Delta a'^{LR\,K^0}_2 \simeq
- {2 g_R\over g_L \lambda}\xi \left(\bar V^R_{u s}+\bar V^{R*}_{c
d} \right) \chi^{K^0} a_2\nonumber\\\label{del2}\eea with $\lambda
= V_{us}$. The direct CP asymmetry in this case can be then
expressed as
\begin{eqnarray}
A^{SM+LR}_{CP} (D^+ \rightarrow  K^0 \pi^+) &=&
{|A^{^{SM+LR}}_{D^+ \rightarrow  K^0 \pi^+}|^2-|\bar
A^{^{SM+LR}}_{D^+ \rightarrow K^0 \pi^+}|^2 \over
|A^{^{SM+LR}}_{D^+ \rightarrow  K^0 \pi^+}|^2+|\bar
A^{^{SM+LR}}_{D^+ \rightarrow K^0 \pi^+}|^2 }= \kappa'\,
\sin(\phi'^{LR}_2-\phi'^{LR}_1)\label{acpLR2}
\end{eqnarray}
where where $\kappa'$ is given as before and the weak phases
$\phi'^{LR}_1$ and $\phi'^{LR}_2$ are defined through \bea
\phi'^{LR}_1 &=& \tan^{-1}\big(\frac{|\Delta a'^{LR }_1|\sin\Delta
\phi'^H_1}{a_1}\big) \nonumber\\\phi'^{LR}_2 &=&
\tan^{-1}\big(\frac{|\Delta a'^{LR \, K^0}_2|\sin\Delta
\phi'^{LR}_2}{a_2}\big)\eea   where $\Delta \phi'^{LR}_1$  and
$\Delta \phi'^{LR}_2$ are the phases of $\Delta a^{LR}_1$ and
$\Delta a^{LR \, K^0}_2$ respectively.

In order to give an estimation of the direct CP asymmetries in
Eqs.(\ref{acpLR1},\ref{acpLR2}) we need to determine the allowed
values of the left right mixing angle  $\xi$ and the elements of
the matrix $\bar V^ R$ relevant to the decay processes under
consideration. Information about the allowed values of the left
right mixing angle $\xi$ can be inferred  from the measurement of
the muon decay parameter $\rho$, which governs the shape of the
overall momentum spectrum, performed by the TWIST collaboration
\cite{MacDonald:2008xf,TWIST:2011aa}. This parameter is related to
$\xi$ via \cite{MacDonald:2008xf}:

\be \rho \simeq \frac{3}{4}\bigg[1- 2\, \big(\frac{g_R}{g_L}
\xi\big)^2\bigg]\ee

Defining $\zeta = \frac{g_R}{g_L} \xi $ and for the TWIST value,
from their latest global fit given in Table VII in
Ref.\cite{TWIST:2011aa}, $\rho = 0.74960 \pm 0.00019$ we obtain
the allowed $2 \sigma$ range of $\zeta$

\be   3.7\times 10^{-3} \lesssim \zeta \lesssim 2.3\times
10^{-2}\label{zeta} \ee

We turn now to discuss the allowed values of the elements of the
matrix $\bar V^R$ appear in Eqs.(\ref{del1},\ref{del2}). These
elements are $ \bar V^R_{ud}, \bar V^R_{cs}, \bar V^R_{us}$ and
$\bar V^R_{cd}$. The real parts of these elements will not produce
any weak CP violating phase required for generating direct CP
asymmetry. In addition, their contributions to the amplitudes will
be always suppressed by a factor $\zeta$ and thus, to a good
approximation, can be neglected compared to the SM contributions.
As a result, we only need to determine the allowed values of the
imaginary parts of $ \bar V^R_{ud}, \bar V^R_{cs}, \bar V^R_{us}$
and $\bar V^R_{cd}$.

 In a recent study, the authors of Ref. \cite{Alioli:2017ces} have
listed the bounds from collider physics, flavor physics, and
low-energy precision measurements on the complex couplings of the
$W^\pm$ boson to right-handed quarks. Particularly, these bounds
are applied to the couplings in the left-right symmetric models
that are generated from the mixing between the charged gauge
bosons of the $S U(2)_R$ and $S U(2)_L$. As shown in Ref.
\cite{Alioli:2017ces}, the experimental value of $(\epsilon '/
\epsilon)_K$ and the stringent bounds on the electric dipole
moment of the neutron can lead to strong bounds on $ Im(\bar
V^R_{ud})$ and $Im(\bar V^R_{us})$. From that study we find that $
Im(\bar V^R_{ud})$ and $Im(\bar V^R_{us})$ can be as large as
$9\times 10^{-4}$ and $2\times 10^{-4}$ respectively. Moreover,
$Im(\bar V^R_{c d})$ can be as large as $2\times 10^{-3}$ without
violating the strongest bounds on the electric dipole moment of
the neutron. The result of the study in Ref.\cite{Alioli:2017ces}
showed also that the dominant constraint on $\zeta Im(\bar V^R_{c
s})$ results from the process $K_L\to \pi^0\, e^+\, e^ -$ and
$\zeta \, Im(\bar V^R_{c s})$ can have a maximum allowed value
$7\times 10^{-3}$. This result shows that we can set  $Im(\bar
V^R_{c s})\simeq {\mathcal O }(1)$ without violating the imposed
constraints. Taking these values into account, we obtain
$|A^{SM+LR}_{CP} (D^+ \rightarrow \bar K^0 \pi^+)|\, \simeq\,
{\mathcal O } (10^{-3})$ which is 8 orders of magnitude larger
than the SM prediction. For the other decay mode we find that
$|A^{SM+LR}_{CP} (D^+ \rightarrow K^0 \pi^+)| \,\simeq \,{\mathcal
O } (10^{-7})$ which is only 4 orders of magnitude larger than the
SM prediction.

\section{Conclusion \label{CONC}}

In this work we have studied CP violation in charged decays of $D$
meson. In particular, we have investigated the direct CP asymmetry
of the Cabibbo favored non-leptonic $D^+ \rightarrow \bar K^0
\pi^+$ and the doubly Cabibbo-suppressed decay mode $D^+
\rightarrow K^0 \pi^+$ within standard model, two Higgs doublet
model with generic Yukawa structure and left right symmetric
models.

In the standard model, we have shown that the generated weak
phases at loop-level are so tiny resulting in  direct CP
asymmetries at the order $10^{-11}$ in both decay modes.

Regarding the two Higgs doublet model with generic Yukawa
structure,after taking into account all relevant constraints onthe
parameter space of the model, we have found that charged Higgs
contributions to the amplitudes can enhance the direct CP
asymmetries 6 and 7 orders of magnitudes with respect to their
standard model predictions for  $D^+ \rightarrow \bar K^0 \pi^+$
and $D^+ \rightarrow K^0 \pi^+$ respectively. Finally, we have
shown that due to the strong constraints on the parameter space of
the LRS models no sizable direct CP asymmetries can be achieved
for the doubly Cabibbo-suppressed decay mode $D^+ \rightarrow K^0
\pi^+$. However, this is not the case for the Cabibbo favored
non-leptonic $D^+ \rightarrow \bar K^0 \pi^+$ decay mode where
sizable direct CP asymmetry of ${\mathcal O } (10^{-3})$ still can
be obtained after respecting all relevant constraints on the
parameter space of the model. This result should motivates search
for direct CP violation in  $D^+ \rightarrow \bar K^0 \pi^+$ decay
mode at colliders. 

\acknowledgments
This work was partially support by CONACYT projects CB-259228 and CB- 286651 and Conacyt-SNI.

\end{document}